\documentclass[%
 reprint,
 superscriptaddress,
 nofootinbib,
 amsmath,amssymb,
 aps,
 prl,
 floatfix,
]{revtex4-1}

\usepackage{graphicx}
\usepackage{dcolumn}
\usepackage{bm}

\begin{document}


\title{Connectivity dynamics in the vitrification of colloidal liquids}

\author{Ruben Higler}
 \affiliation{%
 Physical Chemistry and Soft Matter, Wageningen University and Research, Wageningen, 6708 WE, The Netherlands
 }%
\author{Johannes Krausser}%
\affiliation{%
 Department of Chemical Engineering and Biotechnology,
 and Cavendish Laboratory, University of Cambridge, Cambridge, CB2 3RA, UK
}%
\author{Jasper van der Gucht}
\affiliation{%
 Physical Chemistry and Soft Matter, Wageningen University and Research, Wageningen, 6708 WE, The Netherlands
}%
\author{Alessio Zaccone}
\affiliation{%
 Department of Chemical Engineering and Biotechnology,
 and Cavendish Laboratory, University of Cambridge, Cambridge, CB2 3RA, UK
}%
\author{Joris Sprakel}
\email{joris.sprakel@wur.nl}
\affiliation{%
 Physical Chemistry and Soft Matter, Wageningen University and Research, Wageningen, 6708 WE, The Netherlands
}%

\date{\today}

\begin{abstract}
While various structural and dynamical precursors to vitrification have been identified, a predictive and quantitative description of how subtle changes at the microscopic scale give rise to the steep growth in macroscopic viscosity is missing. It was proposed that the presence of long-lived bonded structures within the liquid may provide this connection. Here we directly observe and quantify the connectivity dynamics in liquids of charged colloids en-route to vitrification. Based on these data, we extend Dyre's elastic model for the glass transition to account for particle-level dynamics; this results in a parameter-free expression for the slowing down of relaxations in the liquid that is in quantitative agreement with our experiments.
\end{abstract}

\maketitle


For fragile glasses, the super-exponential increase in viscosity with small changes in temperature is often described by the phenomenological Vogel-Fulcher-Tamman (VFT) relationship \cite{Angell:1988ie}. The VFT form holds for a wide variety of fragile glass formers, ranging from metallic \cite{Greer:1995iz,Wagner:2011ck} and molecular glasses \cite{Martinez:2001bi,Angell:1995dp} to those formed by polymer chains \cite{Huang:2002ia} or colloidal particles \cite{Brambilla:2009bz}. However, a microscopic interpretation of this universal observation remains elusive. Seminal frameworks for the glass transition, such as mode-coupling theory (MCT), accurately predict the mechanism with which particle motion becomes localised from the static structure alone \cite{Gotze:2012ve}, but cannot recover the VFT law for the viscosity or relaxation time. 

To fully explain and predict vitrification, the emergence of frequency-dependent rigidity must be taken into account. The presence of a finite shear modulus at low frequencies is predicted to underpin the liquid-solid transition also at finite temperature \cite{Trachenko:2013kg,Zhang:2009bv}. This finite-frequency rigidity cannot be understood solely from snapshots of the static structure. Rather, rigidity emerges from long-lived bonds between neighbouring particles \cite{Zaccone:2011cm, Zaccone:2016aa}, which are needed to suppress nonaffine motions characteristic of liquids. It was recently proposed that the same long-lived structures governs their thermodynamics \cite{Trachenko:2013kg}. This implies that long-lived bonded structures may play an important role in the liquid state. The hypothesized connection between such structures and the viscoelasticity of liquids has been verified indirectly, for example for metallic alloys and polymer melts \cite{Krausser:2015vca,LappalaVernon:2016dn}. However, direct microscopic observation of these persistent structures within a liquid is impossible for atomic and molecular systems.

In colloidal glasses, where dynamic arrest can be induced by changing the particle packing fraction, several microstructural and dynamical features have been identified to emerge as the liquid relaxations slow down and the glassy state is approached. These range from heterogeneous dynamics \cite{Weeks:2000dw,Berthier:2011wv,Hedges:2009hb} with features of criticality \cite{Tanaka:2010jo}, localised "soft" vibrational modes \cite{Chen:2011fa} to structural signs in the form of icosahedral order \cite{Leocmach:2012in,Chen:2015ir},  topological clusters \cite{Malins:2013br,Royall:2008fz} and persistent fractal structures \cite{Conrad:2006il}. Nonetheless, a key question remains: Is there a direct and quantitative correlation between the microscopic dynamics of long-lived bonded structures and the slowing down of liquids en-route to kinetic arrest?

In this letter, we study suspensions of charged colloidal particles using three-dimensional confocal microscopy to probe the dynamics of long-lived structures in liquids. Based on these experiments, we reformulate Dyre's elastic model for liquid relaxations. To obtain a prediction for the global relaxation time, we use the dynamics of local coordination number as input to describe the finite-frequency shear modulus within the approach of marginal spring networks. This yields a parameter-free theoretical model that is in quantitative agreement with the experimental data. 

\begin{figure}
\centering
\includegraphics[width=\linewidth]{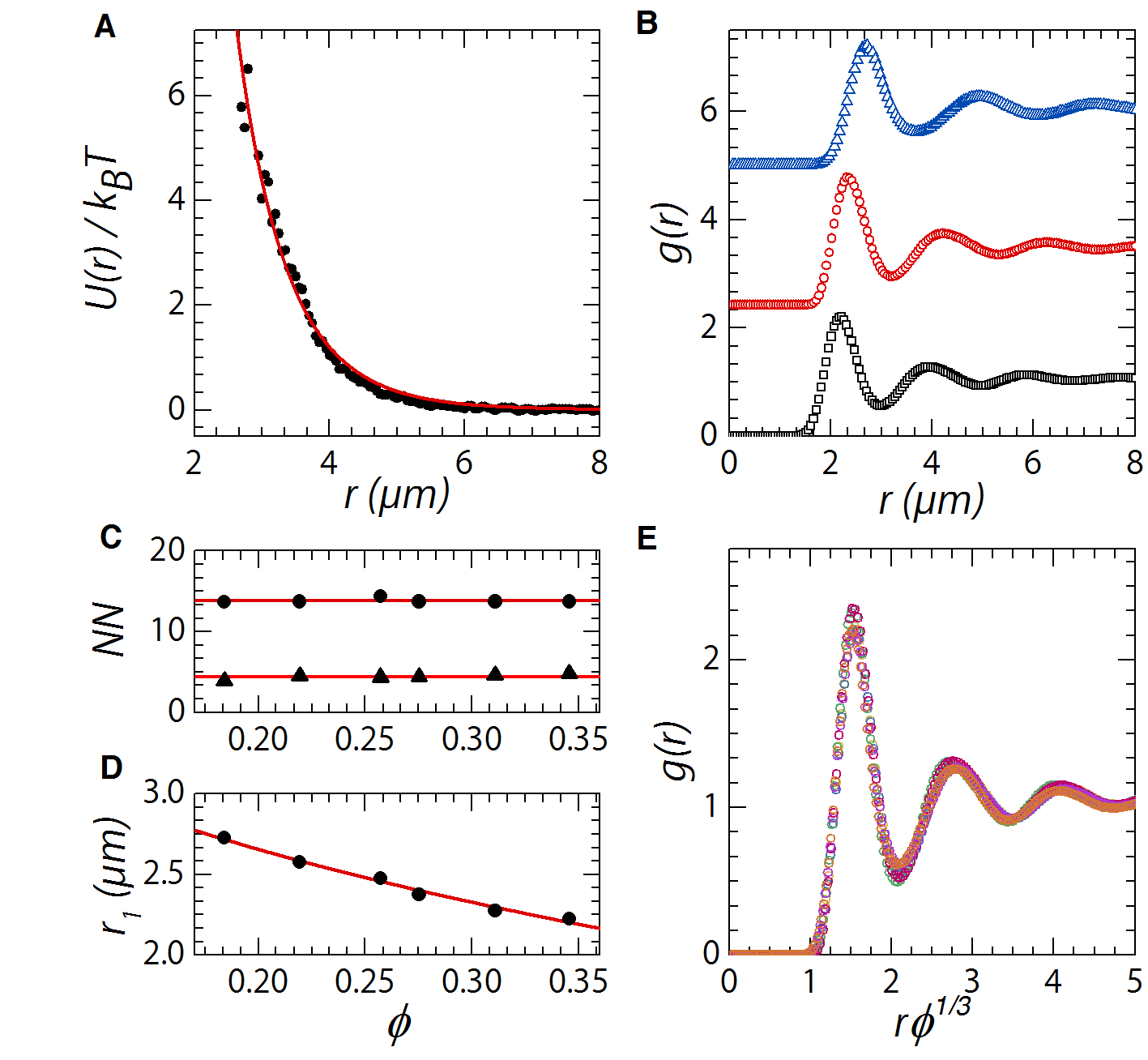} 
\caption{(a) Pair potential determined in experiments (symbols) and fitted to a Yukawa potential (line). (b) $g(r)$ for $\phi = $ (top to bottom) 0.18, 0.28 and 0.35, offset for clarity. (c) $Z(\tau = 0)$ for a neighbor distance of $r_1$, first maximum in $g(r)$ (triangles) or $r_2$ the second minimum in $g(r)$ (circles). (d) Position of the first maximum in $g(r)$, $r_1$. (e) Pair correlation master curve when plotted as a function of $r\phi^{1/3}$, for $\phi = $0.35, 0.31, 0.28, 0.26, 0.22, and 0.18. }
\label{fig:system}
\end{figure}

We study colloidal particles of poly(methyl methacrylate), stabilised by polyhydroxystearic acid, suspended in a density and refractive index matching mixture (see SI). To suppress crystallization we use a mixture of particles with radii $a_{small} = 710$~nm and $a_{large} = 975$~nm. The addition of $10$~mM of the surfactant AOT leads to charging of the particles; in the apolar solvent this results in long-ranged repulsive interactions\cite{Kanai:2015ee,Yethiraj:2003fe}. By inversion of the pair correlation function of a dilute suspension using the hypernetted-chain closure approximation (see SI), we obtain the pair interaction potential (symbols in Fig.~\ref{fig:system}(a)). The experimental data are well-described by the hard-core Yukawa potential $U(r)/k_BT =  \epsilon \frac{\exp{(-\kappa \sigma (\frac{r}{\sigma} - 1))}}{r / \sigma}$ (solid line in Fig.~\ref{fig:system}(a)) with $\sigma = 1.7$ $\mu$m the effective hard sphere diameter, $1/\kappa \approx $ 1.0 $\mu$m the Debye screening length and $\epsilon/k_BT = 28$ the potential at contact.

The instantaneous pair correlation function $g(r,\tau=0)$ displays a liquid-like structure (Fig.~\ref{fig:system}(b)). As the volume fraction of particles is increased, the entire correlation function shifts monotonically to larger values of $r$. This is shown by the shift in the position of the first peak $r_1$ as $r_1 \propto \phi^{-1/3}$, indicative of isotropic compression of the structure (Fig.~\ref{fig:system}(d)). Plotting the $g(r)$ as a function of $r\phi^{1/3}$ yields a collapse of the data within experimental noise (Fig.~\ref{fig:system}(e)). This is corroborated by the fact that the static coordination number $Z(\tau=0)$ from snapshots of the liquid structure is virtually independent of $\phi$ (Fig.~\ref{fig:system}(c)); both when counting neighbors within a distance equal to the first maximum in $g(r)$ ($r_1$) and the first minimum in $g(r)$ ($r_2$).

\begin{figure}
\centering
\includegraphics[width=\linewidth]{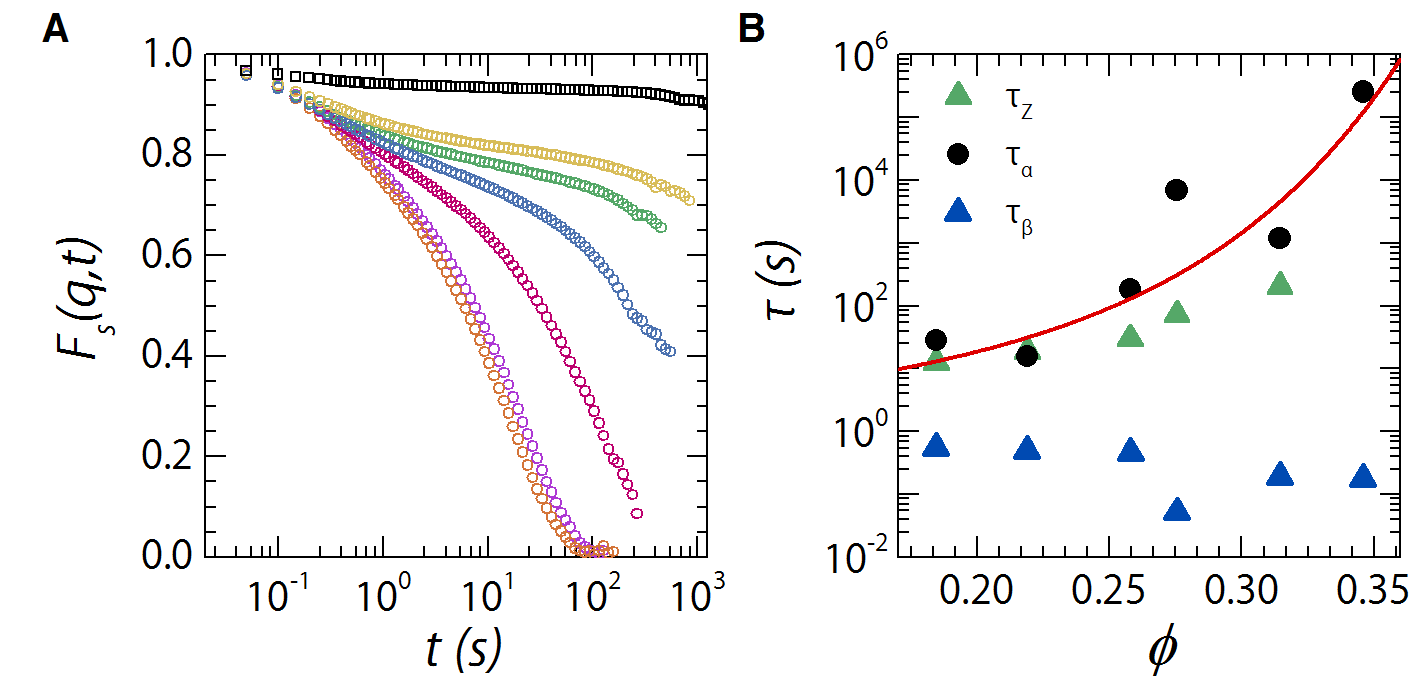}
\caption{Dynamic structure factors and relaxation. (a) Intermediate scattering functions $F_s(q,t)$ at $q = 2\pi/r_1$ for (top to bottom) $\phi = $ 0.49, 0.35, 0.28, 0.31, 0.26, 0.22, and 0.18. (b) Characteristic relaxation times for $\alpha$- (circles) and $\beta$-relaxation processes (triangles). Solid line is our theoretical description, as described in the text.}
\label{fig:slowdown}
\end{figure}

By contrast, over the same range of $\phi$, the particle dynamics change strongly. We compute the intermediate scattering function $F_s(q,t)$ directly from our microscopy data as $F_s(q,t) \allowbreak = \allowbreak \left\langle \allowbreak \exp{( \allowbreak i\textbf{q} \allowbreak \cdot \allowbreak [\textbf{r}(t) \allowbreak - \allowbreak \textbf{r}(0)])} \right\rangle $ with $q = 2 \pi / r_1$. They exhibit two distinct decays (Fig.~\ref{fig:slowdown}(a)); at long lag times structural $\alpha$-relaxation is observed, while local vibrations are seen as a small $\beta$-decay at short times. These are characterised by the relaxation times $\tau_{\alpha}$ and $\tau_{\beta}$, respectively; we extract these two characteristic time scales by fitting the experimental $F_s(q,t)$ to a double stretched-exponential decay\cite{Brambilla:2009bz}.

At high volume fractions, $\phi = 0.49$, we see a time-independent plateau in $F_s(q,t)$ (black squares Fig.~\ref{fig:slowdown}(a)), which indicates full dynamic arrest on experimental time scales. The $\beta$-relaxation time, extracted from $F_s(q,t)$ is virtually independent of $\phi$ and set by the in-cage diffusion coefficient, whereas the structural $\alpha -$relaxation time grows steeply over 5 decades upon approaching the glass transition point $\phi_g$ (Fig.~\ref{fig:slowdown}(b)). 

We observe virtually no changes in the static coordination number, $Z(\tau = 0)$, as a function of particle volume fraction. Over the same range of volume fractions, the particle diffusivity slows down by many orders-of-magnitude. Clearly, the arrest of relaxations in the liquid cannot be explained from the static structure alone. It was recently proposed that the slowing down of relaxation processes in liquids may be understood by considering the formation of persistent bonds between neighboring particles\cite{Trachenko:2013kg,Krausser:2015vca,LappalaVernon:2016dn}

This concept assumes that particles are capable of forming cohesive bonds. Our experimental system is composed of particles interacting with a purely repulsive pair potential  (Fig.~\ref{fig:system}(a)), hence cohesion must be an emergent property caused by many-body correlations. Inversion of the pair-correlation function at finite volume fractions allows us to directly measure the potential-of-mean-force $w(r) = -\ln g(r)$ between the particles. These exhibit a clear bonding minimum at a distance that corresponds to the characteristic nearest-neighbor distance $r_1$. It is these emergent bonds that allows repulsive colloidal systems to build up a finite elastic shear modulus, as demonstrated extensively \cite{Russel:1992vm}. The spring constant $k$ that characterizes the stiffness of these bonds, obtained by fitting the minimum in $w(r)$ to a harmonic well, increases slightly with increasing volume fraction; its absolute value of $k = $3.5-5.5 $k_BT/\mu m^2$ is of order $k_BT/r_1^2$ as expected for a system governed by soft interactions (Fig.~S4). 

To ascertain the dynamics of these emergent bonds, we extract the dynamical coordination number $Z(\tau)$ from our three-dimensional confocal microscopy data. This quantity probes how an initial set of nearest-neighbors changes as time progresses; while the average coordination number at any given snapshot remains the same, particle motion will reshape the cages around a reference particle by breaking existing bonds and reforming new ones such that $Z(\tau)/Z(0) < 1$. 

\begin{figure}
\centering
\includegraphics[width=\linewidth]{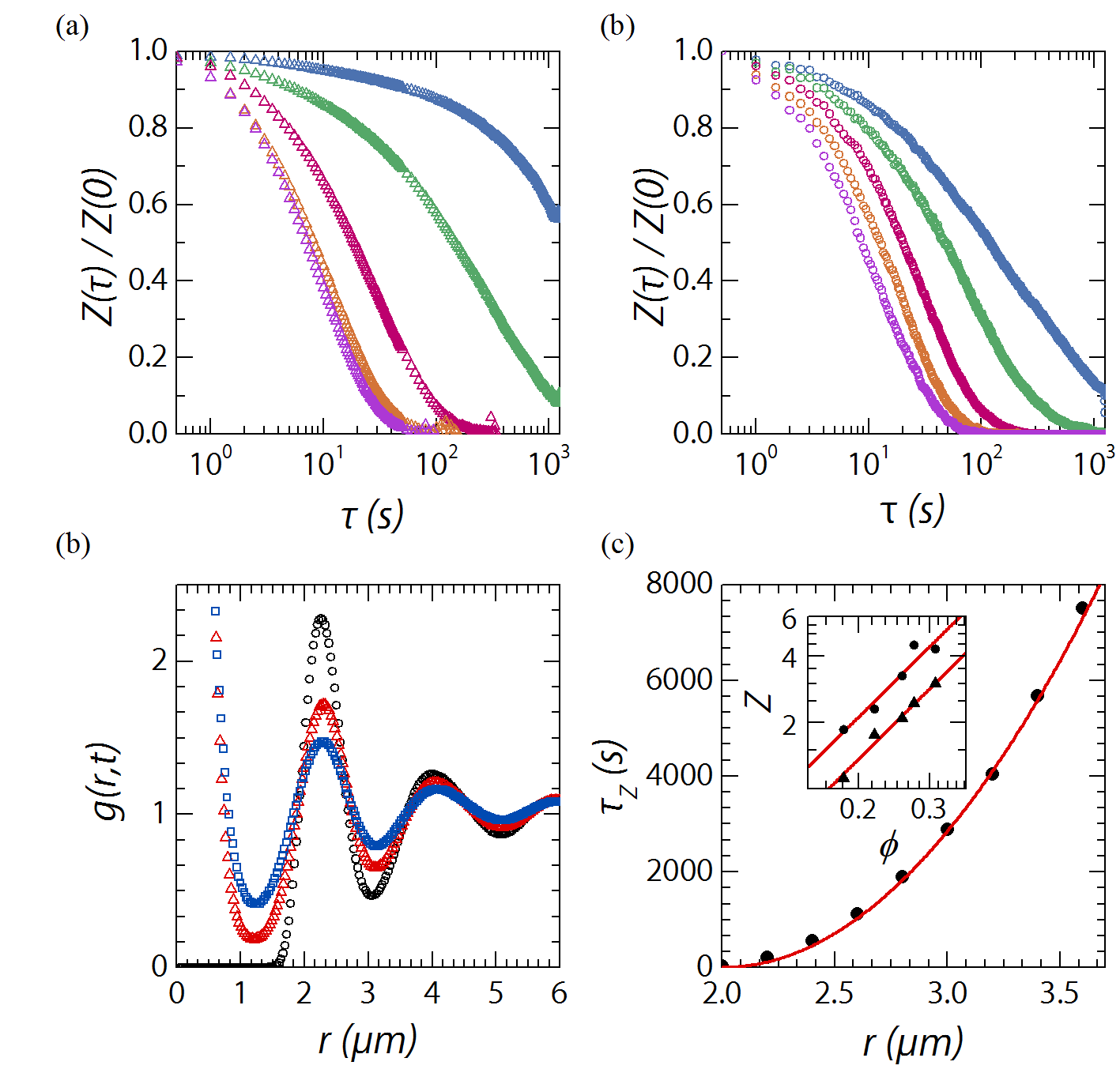} 
\caption{(a) $Z(t)/Z(0)$ from $g(r,t)$, for (left to right) $\phi = $ 0.18, 0.22, 0.26, 0.28, and 0.31. (b) $Z(t)/Z(0)$ as calculated from experimental particle trajectories, for (left to right) $\phi = $ 0.18, 0.22, 0.26, 0.28, and 0.31. (c) $g(r,t)$ for a glassy liquid at $\phi = $0.31 at $t = $0 (black circles), 50 (red triangles), 100 (blue squares) s. (d) Typical decay time of $Z(\tau)/Z(0)$ as a function of cutoff distance $r$. for $\phi = $~0.31. Solid line is a power law fit with an exponent of 2. \textit{inset} Average coordination number $Z$ $\tau_B$ as a function of $\phi$, from van Hove functions (circles) and particle trajectories (triangles). Solid lines are power law fits to the data.}
\label{fig:Zt}
\end{figure}

Experimental particle tracking in three-dimensions never yields a tracking fidelity of 100\%. While we follow most of the particles in our field-of-view for a significant portion of the length of the experiment, interruptions in particle trajectories, will bias the apparent coordination number and its time-evolution. With this in mind, we explore two different approaches to extract $Z(\tau)$ from our experimental data. First we consider the van Hove function
$
g(r,t) = \allowbreak \frac{1}{N}\langle \sum_{i=1}^N \sum_{j=1}^N \allowbreak \delta\left(\textbf{r} + \textbf{r}_i(0) - \allowbreak \textbf{r}_j(t) \right) \rangle
$
where $\delta$ is the Dirac delta function and the brackets indicate an ensemble average. Since we can only track particles for a finite time, it is difficult to discriminate between the self and distinct parts of the van Hove function. Therefore, the computed $g(r,\tau)$ contains both the self- ($i=j$) and distinct ($i \neq j$) parts (Fig.~\ref{fig:Zt}(c))\cite{Weeks:2000dw,Berthier:2011wv}. Our quantity of interest is the distinct part, which probes how cross-correlations between neighbors decay as time progresses (Fig.~\ref{fig:Zt}(c)). From these, the number of bonded neighbours as a function of lag time can be extracted by integration of the first peak as $Z(\tau) = 4 \pi \rho \int_{r_0}^{r_1} \left[ g(r,\tau)-g(r_0,\tau) \right]r^2 dr $, in which $r_0$ and $r_1$ are the first minimum and maximum in $g(r,t)$, respectively and $\rho$ is the particle number density. The term $-g(r_0,\tau)$ in the integral is a first-order correction for contributions of self-correlations to $Z(\tau)$.

The transient nature of a given configuration is reflected in the decay in the $Z(\tau)$ (squares in Fig.~\ref{fig:Zt}(a)). However this approach ignores particle identity; two spatial configurations with equal $Z$ but composed of different particles will incorrectly be ignored as contributing to $Z(\tau)/Z(0)$. Thus, using $g(r,\tau)$ will overestimate the intact number of bonds.

We can also compute $Z(\tau)$ directly from the three-dimensional particle trajectories\cite{Conrad:2006il}. For every particle we determine its neighbors within a distance  $r_1$ at a lag time $\tau$. To obtain the dynamical coordination number we determine the distance between the probe particle, $i$, and its neighbors, $j$, over all lag times, $d_{ij}(\tau) = |\mathbf{r}_i(\tau) - \mathbf{r}_j(\tau)|$. Counting the number of neighbor particles with $d \leq r_1$ at every lag time gives us the dynamical coordination number $Z_i(\tau)$(Fig.~\ref{fig:Zt}(b)).

With this method, unphysical bond breaking events will be detected when a trajectory is erroneously truncated, and a new one generated, if a particle is missed during the locating procedure even for a single frame. In such an event, the particle stays in place physically, but it is assigned a new identity. While this particular experimental artifact occurs rarely, it cannot be completely avoided. As such, $Z(\tau)$ computed from particle trajectories will underestimate the number of intact bonds; indeed, differences emerge between the two approaches to compute the dynamic coordination number (Fig.~\ref{fig:Zt}(a) and (b)). This difference seems to be largest for samples at higher volume fractions, while for purely liquid samples the two methods give similar decay curves. 

A crucial parameter in this determination of the dynamic coordination number is the neighbor cut-off distance $r$. To test the effect of this characteristic length scale on the resulting $Z(\tau)$, we determine the decay time $\tau_Z$ of $Z(\tau)/Z(0)$, by fitting $Z(\tau)/Z(0)$ with a stretched exponential, as a function of $r$, for $\phi = 0.31$ (symbols in Fig.~\ref{fig:Zt}(d)). We find $\tau_Z \propto r^2$ (solid line Fig.~\ref{fig:Zt}(d)), indicating the diffusive nature of neighbour exchange as expected in the liquid state.

It is interesting to note that the nearest-neighbor exchange dynamics probed by $Z(\tau)$ is a different measure for relaxations in the liquid than the single-particle mobility probed in $F_s(q,t)$. The dynamical coordination number probes how particles move with respect to its bonded neighbors. For example, the sliding of two particles with respect to eachother, while remaining bonded, or the collective plug-like motion of a cluster of particles in a shear-transformation zone,  does not lead to a reduction in $Z$ but does lead to a decorrelation of $F_s(q,t)$. By contrast, cage rattling may break bonds such that $Z(\tau)$ decays while it resulting in only very weak decay in the dynamic structure factor. Indeed, the characteristic timescale for reconfiguration of a coordination shell $\tau_Z$ is lower and grows less steeply than $\tau_{\alpha}$  obtained from fitting $F_s(q,t)$ (Fig.~\ref{fig:slowdown}(b)).

Of special interest is the scaling of $Z(\tau)$ as a function of $\phi$ as this provides a clue to the effect of local coordination dynamics on the volume fraction induced quenching of relaxation processes. While the absolute magnitude of  $Z(\tau, \phi)$ differs between our two different methods, their scaling exponent with $\phi$ is independent of the method (Fig.~\ref{fig:Zt}(d) \textit{inset}).

\begin{figure}
\centering
\includegraphics[width=\linewidth]{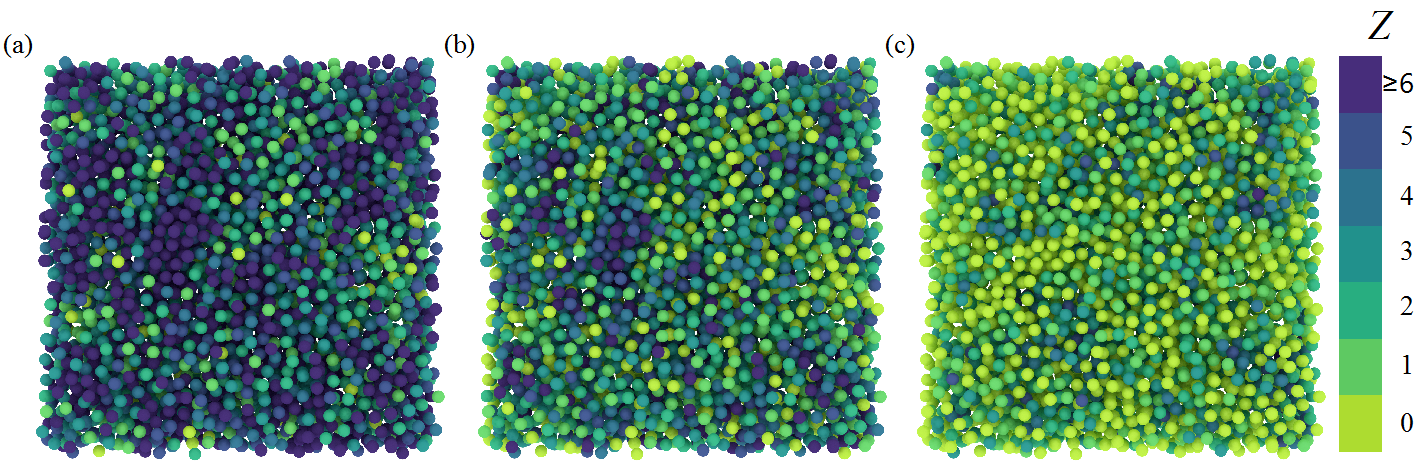} 
\caption{(a-c) computer-generated renderings of our experimental system, in which the particles are color-coded according to their actual coordination number.(a): $Z(\tau=0)$, (b): $Z(\tau = \tau_B)$, and (c): $Z(\tau = 250~s)$. Dark blue particles have $Z\geq Z_c$, while particles with $Z<6$ are colored in increasing shades of yellow, shown for $\phi =$~0.31.}
\label{fig:renders}
\end{figure}

A prototypical feature of liquids approaching their glass transition is that their dynamics becomes heterogeneous\cite{Weeks:2000dw,Berthier:2011wv,Hedges:2009hb}. To probe the spatial homogeneity of coordination structures, we reconstruct our experimental data by color-coding particles according to their coordination number $Z$, taken both as the static structure $Z(\tau=0)$, at the Brownian timescale $Z(t=\tau_B \sim 10^1 s)$, and at long timescales $Z(\tau = 250~s)$. Indeed, we observe not only how the average coordination number decreases as the volume fraction is reduced, but also how the distribution of coordination numbers is strongly heterogeneous in space (Fig.~\ref{fig:renders}). From the reconstructions it is also clear that the debonding events through which the sample loses rigidity do not occur homogeneously throughout the sample; areas of high connectivity stay connected while areas with low connectivity weaken rapidly. 

In order for local structure to contribute to rigidity, the cage that surrounds a central particle, needs to be intact for a least as long as the required time of escape. Bonds that break before the attempted escape from a cage, do not contribute to the slowing down of particle dynamics. The characteristic timescale of particle escape from a cage will be of the order of the Brownian time scale $\tau_B = a^2/D \sim 10^1$~s, with $D$  the particle self-diffusion coefficient. Thus, bonds which live longer than $\tau_B$ can contribute to stable interconnected structure.  
The concept that long-lived neighbors contribute to the formation of rigid structure in the liquid, revolves around the idea that the transition from a liquid-like to a solid-like response is signalled by the formation of an isostatic structure of  load-bearing bonds at a characteristic frequency. Note that this is not the same as the zero-frequency liquid-solid transition, which is the focus of the jamming framework, that signals the arrest of flow on all timescales. Rather, the location of this liquid-solid transition in these thermal fluids will depend on the choice of frequency; for the purposes of this discussion we use the characteristic frequency $\omega_B = 1/\tau_B$. 

From the experimental data for $Z(\tau)$, we can measure the value of the coordination number at $\tau = \tau_B$ as a proxy for the amount of bonds that could contribute to rigidity (inset Fig.~\ref{fig:Zt}(d)). We find that $Z(\tau_B)$ grows as the particle volume fraction is increased, whereas the static coordination number $Z(\tau=0)$ remains constant over the same range of volume fractions (Fig.~\ref{fig:system}(c)).

For the frequency of interest, $\omega_B$, a liquid-solid transition must emerge at some critical volume fraction $\phi_c$, where in accord with the Maxwell criterion for isostaticity in three-dimensional central-force lattices, $Z(\tau_B) = Z_c = 6$. We find that our experimental data, over the limited range of volume fractions accessible in our experiments, can be well described by $Z(\tau_B) = Z_c(\frac{\phi}{\phi_c})^b$, with an exponent $b = 1.8$. This leads us to identify the transition between a liquid and a solid like response at a frequency $\omega_B$ at $\phi_c \approx 0.45$.

To explain these data, we adopt the phenomenological model developed by Dyre. This approach treats relaxation events, as localized shear deformations of the surrounding medium. By combining thermally-activated dynamics of the Eyring type with continuum mechanics, Dyre analytically derived a relationship between the structural relaxation time and the shear modulus $G$ of the liquid \cite{Dyre:1998jm} as $\tau_{\alpha} = \tau_B \exp\left[ {\frac{G V_a}{k_BT}}\right]$, 
where $V_a$ is the activation volume. The term $E_a=GV_a$ represents the elastic activation energy to expand the cage allowing for irreversible rearrangements. It is important to note that in Dyre's model, the local shear rigidity is described by the high-frequency shear modulus, which is affine by definition. 

A microscopic interpretation of the shear modulus is provided by the framework for disordered bead-spring lattices \cite{Zaccone:2013gl}, which gives the affine shear modulus as $G = \frac{1}{5\pi} \frac{k(\phi)}{r_1(\phi)}\phi Z$\cite{Zaccone:2011cm}, with $k(\phi)$ the effective spring constant of the interparticle bonds(inset Fig.~S4) and $r_1(\phi)$ the average interparticle spacing.

We estimate the activation volume $V_a$ as the cage volume $V_a = \frac{4}{3}\pi r_1(\phi)^3$. We now obtain a microscopic version of Dyre's model: $\tau_{\alpha} = \tau_B \exp\left[{\frac{4 k(\phi) r_1(\phi)^2 Z_c \phi^{1+b} \phi_c^{-b}}{15 k_BT}}\right]$ in which all microscopic parameters can be directly extracted from the experiments. Without any adjustable parameters, this model provides a quantitative agreement with experimentally determined values for the relaxation time $\tau_{\alpha}$ (solid line in Fig.~\ref{fig:slowdown}(b)). 

The excellent agreement between experimental data and theory illustrates how liquid viscosity at the global scale can be understood from the existence and dynamics of emerging bonds between neighboring particles. This suggests that the glass transition in these systems, in analogy to vitrification in metallic alloys or polymer melts \cite{Krausser:2015vca,LappalaVernon:2016dn}, is a dynamical connectivity transition leading to an isostatic condition at a finite, relevant, frequency.

We thank Ties van de Laar, Remco Fokkink and Raoul Frijns for technical assistance with particle synthesis and characterisation. 

%

\end{document}